\documentclass[a4paper,fleqn,usenatbib]{mnras}

\usepackage{newtxtext,newtxmath}

\usepackage[T1]{fontenc}
\usepackage{ae,aecompl}

\usepackage{graphicx}	
\usepackage{amsmath}	
\usepackage{amssymb}	
\newcommand\T{\rule{0pt}{1.6ex}}  
\usepackage[export]{adjustbox}
\usepackage{float}
\usepackage{stfloats}
\usepackage{tabularx}
\usepackage{hyperref}
\usepackage{subcaption}
\usepackage{lineno}

\title[MAGIC and \textit{Fermi}-LAT results on HAWC sources]{MAGIC and \textit{Fermi}-LAT gamma-ray results on unassociated HAWC sources}

\author[The MAGIC, HAWC and \textit{Fermi}-LAT Collaborations]{
\parbox{\textwidth}{
\normalsize
M.~L.~Ahnen$^{1}$,
S.~Ansoldi$^{2,20}$,
L.~A.~Antonelli$^{3}$,
C.~Arcaro$^{4}$,
D.~Baack$^{5}$,
A.~Babi\'c$^{6}$,
B.~Banerjee$^{7}$,
P.~Bangale$^{8}$,
U.~Barres de Almeida$^{8,9}$,
J.~A.~Barrio$^{10}$
% ...
(MAGIC Collaboration),
A.~Albert$^{26}$,
R.~Alfaro$^{27}$,
C.~Alvarez$^{28}$,
R.~Arceo$^{28}$,
J.~C.~Arteaga-Vel\'{a}zquez$^{29}$,
D.~Avila~Rojas$^{27}$,
H.A.~Ayala~Solares$^{30}$,
A.~Becerril$^{27}$,
E.~Belmont-Moreno$^{27}$,
S.~Y.~BenZvi$^{31}$
% ...
(HAWC Collaboration),
M.~Ajello$^{56}$ 
L.~Baldini$^{57}$ 
G.~Barbiellini$^{58,59}$ 
B.~Berenji$^{60}$ 
E.~Bissaldi$^{61,62}$ 
R.~D.~Blandford$^{63}$ 
R.~Bonino$^{64,65}$ 
E.~Bottacini$^{63,66}$ 
T.~J.~Brandt$^{67}$ 
J.~Bregeon$^{68}$ 
P.~Bruel$^{69}$ 
R.~A.~Cameron$^{63}$ 
R.~Caputo$^{70}$ 
P.~A.~Caraveo$^{71}$ 
D.~Castro$^{72,67}$ 
E.~Cavazzuti$^{73}$ 
G.~Chiaro$^{71}$ 
S.~Ciprini$^{74,75}$ 
D.~Costantin$^{76}$ 
F.~D'Ammando$^{77,78}$ 
F.~de~Palma$^{62,79}$ 
A.~Desai$^{56}$ 
N.~Di~Lalla$^{57}$ 
M.~Di~Mauro$^{63}$ 
L.~Di~Venere$^{61,62}$ 
A.~Dom\'inguez$^{80}$ 
C.~Favuzzi$^{61,62}$ 
Y.~Fukazawa$^{81}$ 
S.~Funk$^{82}$ 
P.~Fusco$^{61,62}$ 
F.~Gargano$^{62}$ 
D.~Gasparrini$^{74,75}$ 
N.~Giglietto$^{61,62}$ 
F.~Giordano$^{61,62}$ 
M.~Giroletti$^{77}$ 
T.~Glanzman$^{63}$ 
D.~Green$^{83,67}$ 
I.~A.~Grenier$^{84}$ 
S.~Guiriec$^{85,67}$ 
A.~K.~Harding$^{67}$ 
E.~Hays$^{67}$ 
J.W.~Hewitt$^{86}$\thanks{\href{john.w.hewitt@unf.edu }{john.w.hewitt@unf.edu}}
D.~Horan$^{69}$ 
G.~J\'ohannesson$^{87,88}$ 
M.~Kuss$^{89}$ 
S.~Larsson$^{90,91}$ 
I.~Liodakis$^{63}$ 
F.~Longo$^{58,59}$ 
F.~Loparco$^{61,62}$ 
P.~Lubrano$^{75}$ 
J.~D.~Magill$^{83}$ 
S.~Maldera$^{64}$ 
A.~Manfreda$^{57}$ 
M.~N.~Mazziotta$^{62}$ 
I.Mereu$^{92}$ 
P.~F.~Michelson$^{63}$ 
T.~Mizuno$^{93}$ 
M.~E.~Monzani$^{63}$ 
A.~Morselli$^{94}$ 
I.~V.~Moskalenko$^{63}$ 
M.~Negro$^{64,65}$ 
E.~Nuss$^{68}$ 
N.~Omodei$^{63}$ 
M.~Orienti$^{77}$ 
E.~Orlando$^{63}$ 
J.~F.~Ormes$^{95}$ 
M.~Palatiello$^{58,59}$ 
V.~S.~Paliya$^{56}$ 
M.~Persic$^{58,96}$ 
M.~Pesce-Rollins$^{89}$ 
V.~Petrosian$^{63}$ 
F.~Piron$^{68}$ 
T.~A.~Porter$^{63}$ 
G.~Principe$^{82}$ 
S.~Rain\`o$^{61,62}$ 
B.~Rani$^{67}$ 
M.~Razzano$^{89,97}$ 
S.~Razzaque$^{98}$ 
A.~Reimer$^{99,63}$ 
O.~Reimer$^{99,63}$ 
C.~Sgr\`o$^{89}$ 
E.~J.~Siskind$^{100}$ 
G.~Spandre$^{89}$ 
P.~Spinelli$^{61,62}$ 
H.~Tajima$^{101,63}$ 
M.~Takahashi$^{102}$ 
J.~B.~Thayer$^{63}$ 
D.~J.~Thompson$^{67}$ 
D.~F.~Torres$^{103,104}$ 
E.~Torresi$^{105}$ 
E.~Troja$^{67,83}$ 
J.~Valverde$^{69}$ 
K.~Wood$^{106}$ 
M.~Yassine$^{58,59}$
(Fermi-LAT Collaboration),
}\newauthor
(Affiliations can be found after the references)
}

\date{Accepted XXX. Received YYY; in original form ZZZ}

\pubyear{2017}

\begin{document}
\label{firstpage}
\pagerange{\pageref{firstpage}--\pageref{lastpage}}
\maketitle
\clearpage
\begin{abstract}
The HAWC Collaboration released the 2HWC catalog of TeV sources, in which 19 show no association with any known high-energy (HE; $E\gtrsim 10$ GeV) or very-high-energy (VHE; $E\gtrsim 300$ GeV) sources. This catalog motivated follow-up studies by both the MAGIC and \textit{Fermi}-LAT observatories with the aim of investigating gamma-ray emission over a broad energy band. In this paper, we report the results from the first joint work between HAWC, MAGIC and \textit{Fermi}-LAT on three unassociated HAWC sources: 2HWC J2006+341, 2HWC J1907+084$^{*}$ and 2HWC J1852+013$^{*}$. Although no significant detection was found in the HE and VHE regimes, this investigation shows that a minimum $1\degr$ extension (at 95\% confidence level) and harder spectrum in the GeV than the one extrapolated from HAWC results are required in the case of 2HWC J1852+013$^{*}$, while a simply minimum extension of $0.16\degr$ (at 95\% confidence level) can already explain the scenario proposed by HAWC for the remaining sources. Moreover, the hypothesis that these sources are pulsar wind nebulae is also investigated in detail. 
\end{abstract}

\begin{keywords}
gamma rays: general -- gamma rays: individual: (2HWC J2006+341, 2HWC J1907+084$^{*}$, 2HWC J1852+013$^{*}$) -- pulsars: general
\end{keywords}

\section{Introduction}
The synergy of current gamma-ray observatories is a unique opportunity to explore the Universe between a few tens of MeV to hundreds of TeV. The study of such a broad energy band can be accomplished by means of different detection techniques.  At the highest energies are the Water Cherenkov arrays, such as the High Altitude Water Cherenkov (HAWC) Observatory, which is sensitive to cosmic and gamma rays between 100 GeV and 100 TeV. A large effective area and high duty cycle make HAWC an optimal instrument to perform survey studies on multi-TeV sources. Other detection techniques acheive a better sensitivity than HAWC at energies below approximately 10 TeV: the Imaging Atmospheric Cherenkov Telescope (IACT) technique, as implemented by the Major Atmospheric Gamma-ray Imaging Cherenkov (MAGIC) telescopes, provides better angular and energy resolution. However, the duty cycle of such atmospheric Cherenkov telescopes is greatly reduced by  high-intensity background light or non-optimal weather conditions at night. At even lower energies ($\sim$ GeV), detectors on board satellites, like the \textit{Fermi}-Large Area Telescope (LAT), are used to detect gamma rays. This instrument has a high duty cycle, a wide field of view (FoV) of 20\% of the sky and very good gamma/hadron separation. \\

\begin{table*}
\begin{center}
\caption{
Coordinates, photon index, flux at the pivot energy (7 TeV) and energy range for the analyzed sources. Values are provided in the 2HWC catalog, except for the energy range, which was obtained in a dedicated analysis. Only statistical uncertainties are shown. Based on a study of the Crab Nebula by HAWC \citep{CrabHAWC}, the systematic uncertainty can be divided into several components: 0.10$\degr$ in angular resolution, 0.2 in photon index, and 50\% in flux normalization.}
\label{HAWC_parameters}      
	\begin{tabular}{ c | c c c c c c c c |}
		\cline{2-9}
		& RA & Dec & l & b &1$\sigma$ stat. error & Photon index& Flux normalization & Energy range\\
		& [$^{\circ}$] & [$^{\circ}$] & [$^{\circ}$] & [$^{\circ}$] & [$^{\circ}$]  &   & [$\times10^{-15}$ TeV$^{-1}$cm$^{-2}$s$^{-1}$] & [TeV]\\
		\hline
	     \multicolumn{1}{| c | }{2HWC J2006+341} & 301.55 & 34.18 & 71.33 & 1.16 & 0.13 & $2.64\pm0.15$ & $9.6\pm1.9$ & 1 -- 86 \\
		\hline
	     \multicolumn{1}{| c |}{2HWC J1907+084$^{*}$} & 286.79 & 8.50 & 42.28 & 0.14 & 0.27 & $3.25\pm0.18$ & $7.3\pm2.5$ & 0.18 -- 10 \\
		\hline
	     \multicolumn{1}{| c |}{2HWC J1852+013$^{*}$} & 283.01 & 1.38 & 34.23 & 0.50 & 0.13 & $2.90\pm0.10$ & $18.2\pm2.3$ & 0.4 -- 50\\
		\hline	
	\end{tabular}
	\end{center}
\end{table*}

HAWC has published two catalogs of TeV sources: 1HWC for sources in the inner Galactic plane using 275 days of data with a configuration of approximately one-third of the full array (HAWC-111; \citealt{2016ApJ...817....3A}), and 2HWC for almost the entire sky using 507 days of the completed HAWC detector \citep{2HWC_Catalog}. The second catalog improves over the first with respect to exposure time, detector size and angular resolution, resulting in a significant improvement in sensitivity. As done for the previous catalog, 2HWC data was analyzed using a binned likelihood method described in \cite{2015arXiv150807479Y}. In this method a source model needs to be assumed for all sources in the sky. The model for each source is characterized by the source morphology and its spectrum. For the 2HWC analysis, HAWC used two different approaches: (1) a point-like search adopting a spectrum defined by a power-law function, d$N$/d$E= N_{0} \left(\frac{E}{E_{0}}\right)^{-\Gamma}$ (with $N_{0}$ the normalization, $E_{0}$ the pivot energy and $\Gamma$ the spectral index), with spectral index $\Gamma=2.7$, and (2) extended source searches with a source morphology modeled as a uniform disk of $0.5\degr$, $1\degr$ and $2\degr$ in radius and spectral index $\Gamma=2.0$. The total number of sources identified in this catalog was 39, of which 19 were not associated with any previously reported TeV source within an angular distance of $0.5\degr$. All 2HWC sources presented a test statistic ($TS$) above $25$ (equivalent to a pre-trial significance of $\sim5\sigma$).  

The 2HWC catalog motivated follow-up studies with H.E.S.S. \citep{HESSHAWC}, VERITAS \citep{VERITASHAWC} and also MAGIC and \textit{Fermi}-LAT. In this work, we focused on the 19 sources with no high-energy (HE; $E\gtrsim 10$ GeV) or very-high-energy (VHE; $E\gtrsim 300$ GeV) association, in order to provide new multi-wavelength information of candidates without a lower energy counterpart. After evaluating those sources, a short list of three targets was selected: 
2HWC J2006+341 ($\textrm{RA}=301.55\degr$, $\textrm{Dec}=34.18\degr$), 2HWC J1907+084$^{*}$ ($\textrm{RA}=286.79\degr$, $\textrm{Dec}=8.50\degr$) and 2HWC J1852+013$^{*}$ ($\textrm{RA}=283.01\degr$, $\textrm{Dec}=1.38\degr$). These sources were chosen because they lie in the FoV of previous MAGIC observations, allowing MAGIC to analyze these sources without performing new dedicated observations.  

Even though the HAWC spectra of each source were determined using a likelihood fit, the 2HWC catalog did not use a likelihood method to describe multiple sources simultaneously. In the 2HWC catalog, the asterisk of 2HWC J1907+084* and 2HWC J1852+013* indicates that the sources were near another source with larger significance and thus their characterization may be influenced by neighboring sources.
2HWC J2006+340, 2HWC J1907+084* and 2HWC J1852+013* were detected in the point source search with significances of $6.10\sigma$, $5.80\sigma$ and $8.50\sigma$, respectively. The corresponding photon index and flux normalization values obtained in the 2HWC catalog are listed in \autoref{HAWC_parameters}. Their corresponding energy range is computed with a dedicated HAWC analysis (see Section \ref{HAWCinfo}) and also given in the table.

\begin{table*}
\begin{center}
\caption{Distance in degrees between the four MAGIC wobble pointing positions (W1, W2, W3 and W4) and the selected 2HWC sources. The total time after data quality cuts, in hours, achieved by MAGIC in each case is also shown.} 
\label{table:dist_wobblepos}      
	\begin{tabular}{ c | c  c | c c | c c |c c |}
		\cline{2-9}
		& \multicolumn{2}{c}{W1} & \multicolumn{2}{c}{W2} & \multicolumn{2}{c}{W3} & \multicolumn{2}{c |}{W4}\\
		\cline{2-9}
		& Distance [$\degr$] & t$_{\textrm{total}}$ [h] & Distance [$\degr$] & t$_{\textrm{total}}$ [h] & Distance [$\degr$] & t$_{\textrm{total}}$ [h] & Distance [$\degr$] & t$_{\textrm{total}}$ [h] \\
		\hline
	     \multicolumn{1}{| c | }{2HWC J2006+341} & 0.5 & 16.0 & 0.9 & 14.0 & 0.4 & 16.3 & 1.0 & 14.8\\
		\hline
	     \multicolumn{1}{| c |}{2HWC J1907+084$^{*}$} & 0.5 & 1.0 & 1.2 & 1.0 & 0.7 & 1.3 & 1.1 & 0.9\\
		\hline
	     \multicolumn{1}{| c |}{2HWC J1852+013$^{*}$} & 1.1 & 30.8 & 0.7 & 28.8 & 1.2 & 29.6 & 0.6 & 27.5\\
		\hline	
	\end{tabular}
	\end{center}
\end{table*}

The paper is structured as follows: in Section \ref{section2}, the data analysis for MAGIC and \textit{Fermi}-LAT observations are presented. Description of the specific HAWC analysis on the selected sources is also included. The observations and results, for each source separately, are shown in Section \ref{section3}. Discussion and conclusion can be found in Sections \ref{section4} and \ref{sec:conclusions}, respectively. 

\section{Data Analysis}
\label{section2}

\subsection{HAWC}
\label{HAWCinfo}
The HAWC Observatory is the second generation of ground-based gamma-ray extensive air shower arrays, located in Sierra Negra, Mexico ($19.0\degr$N, $97.3\degr$ W, 4100 m a.s.l.), and successor to the Milagro Gamma-ray Observatory. The current system, inaugurated on March 20 2015, is comprised of 300 water Cherenkov detectors (WCD) over an area of 22,000 m$^{2}$. Science operations began before detector completion, under the HAWC-111 configuration. The angular resolution of HAWC varies with event size (fraction of photo-multiplier tubes reporting a signal or $f_{hit}$) from $0.17\degr$ to $1.0\degr$ \citep{CrabHAWC}. HAWC operates with $>95$\% duty cycle with a large FoV of 15\% of the sky, which allows it to scan two-thirds of the sky every 24 hours.

Information presented here on 2HWC J2006+341, 2HWC J1907+084* and 2HWC J1852+013* is taken mostly from the 2HWC catalog \cite{2HWC_Catalog}. The only exception is the energy range shown in \autoref{HAWC_parameters}. The likelihood analysis in HAWC is computed over $f_{hit}$ bins, which can be considered an energy estimator. However, the $f_{hit}$ bins depend strongly on the declination and spectral hardness of each source, and so does this $f_{hit}$/energy correlation. The energy range is then given as a constraint on the photon distribution as a function of $f_{hit}$ for each separate source. Following \cite{2HWC_Catalog}, we take the energy range as the boundaries within which the events contribute to the 75\% of the $TS$ value. 

\subsection{MAGIC}
\label{MAGICinfo}
MAGIC is a stereoscopic system of two 17 m diameter IACTs situated on the Canary island of La Palma, Spain ($28.8\degr$N, $17.8\degr$ W, 2225 m a.s.l.). The current system achieves an integral sensitivity of $0.66\pm0.03$\% of the Crab Nebula flux (CU) in 50 hours of observation above 220 GeV \citep{2016APh....72...76A}. The energy threshold in stereoscopic mode is as low as 50 GeV at low zenith angles under dark observational conditions \citep{2012APh....35..435A}. 

The analysis presented in this work is performed using the standard MAGIC analysis software (MARS; \citealt{Zanin2013}).  The significance is computed following Eq. 17 of \cite{LiMa}. Differential and integral flux upper limits (ULs) are calculated using the Rolke algorithm \citep{Rolke2005} with a confidence level (CL) of 95\%, assuming a Poissonian background and a total systematic uncertainty of 30\%.

As mentioned above, the three analyzed sources were included in the FoV of former MAGIC observations. These archival data were taken using the false-source tracking mode, or \textit{wobble-mode}: the telescopes point at four different positions located $0.4\degr$ from the nominal source, which allows us to evaluate the background simultaneously \citep{Fomin1994}. Thus, our observations were not dedicated to the 2HWC sources and so, their coordinates are shifted from the camera center by different distances than the standard offset of $0.4\degr$  (see \autoref{fig:skymap}). To account for their location in the camera, the background used in the calculation of ULs was evaluated through the \textit{off-from-wobble-partner} (OfWP) method \citep{Zanin2013}. \autoref{table:dist_wobblepos} summarizes the distances between the camera center and the 2HWC sources at the four different \textit{wobble} positions. The total observation time, after data quality cuts, for each case is also quoted in \autoref{table:dist_wobblepos}. It is worth highlighting that the MAGIC sensitivity depends on the angular offset from the pointing direction. However, after the MAGIC upgrade of 2011--2012, the sensitivity at offset angles larger than $0.4\degr$ improved considerably as shown by \cite{2016APh....72...76A}. For the analysis performed in this work, and given the range of angular offsets for all the candidates, the sensitivity remains between $\sim 0.6-1.0$ \% CU. 

Observations of 2HWC J1852+013$^{*}$ were  carried out entirely under dark conditions, i.e. in absence of moonlight. On the other hand, 2HWC J2006+341 and 2HWC J1907+084$^{*}$ were observed with nominal high-voltage at background levels ranging between 1 and 8 times the brightness of the dark sky due to different Moon phases. The higher the moonlight level, the brighter is the night sky background and therefore, stronger cuts to the signal are applied during this analysis, following the prescription of \cite{MAGICMoonPerformance}. This is taken into account by selecting appropriate Monte Carlo-simulated gamma-ray and background data to match the observational conditions. The background data are used for the computation of the gamma/hadron separation through the Random Forest (RF), a multi-dimensional classification algorithm based on decision trees \citep{RFMAGIC}.

Since the three HAWC sources each have a maximum significance in the point-source HAWC maps, they may be point-like sources for MAGIC as well. Therefore, we analyze the candidates under two hypotheses: we assume that the sources are point-like for MAGIC (point spread function; PSF $\lesssim0.10\degr$, beyond a few hundred GeV) or are extended with a radius of $0.16\degr$. Larger extensions cannot be adopted due to the OfWP method and the standard 0.4$^{\circ}$ offset applied in the \textit{wobble} pointing mode, because the expected region of gamma-ray emission from the 2HWC source and the background regions selected to compute flux ULs would overlap.\\

\subsection{\textit{Fermi}-LAT}
\label{Fermiinfo}
The LAT on board the \textit{Fermi Gamma-Ray Space Telescope} has continuously monitored the sky since 2008. It is sensitive to HE gamma rays between 20 MeV and $\sim$1 TeV \citep{2009ApJ...697.1071A} and scans the entire sky every 3 hours. For this work we used data taken between 2008 August and 2017 February based on the \texttt{Pass 8} SOURCE photon reconstruction. The \texttt{Pass 8} data offer two primary benefits for the study of HE gamma-ray sources: a greater acceptance compared with previous LAT reconstructions and an improved PSF with a 68\% containment angle less than 0.2$\degr$ above 10 GeV that is nearly constant with increasing energy \citep{2013arXiv1303.3514A}.

For each source of interest we analyze energies between 10 GeV and 1 TeV using the standard binned likelihood framework provided by the Fermi Science Tools (v10r01p01). Data within a 10$^{\circ}$ radius of interest were binned into 8 energy bins per decade and a spatial bin size of 0.05$^{\circ}$. We used the recommended Galactic and isotropic backgrounds\footnote{Galactic interstellar emission model: $gll\_iem\_v06.fits$, Isotropic:  $iso\_P8R2\_SOURCE\_V6\_v06.txt$. Please see: \url{http://fermi.gsfc.nasa.gov/ssc/data/access/lat/BackgroundModels.html}}.

We did not extend our analysis to lower energies for two primary reasons: for any HAWC source to be detected at lower energies, it must be detectable at $>$10 GeV with the LAT, unless the HAWC and LAT emission is produced by a different component; and this high-energy cut suppresses photons from gamma-ray pulsars and Galactic diffuse emission in the plane. As a source model for this analysis we use the Third Catalog of Hard \textit{Fermi}-LAT Sources for point sources \citep[3FHL;][]{2017ApJS..232...18A} and the Fermi Galactic Extended Source (FGES) catalog for extended sources \citep{2017ApJ...843..139A}.

Using $\sim$8.5 years of \texttt{Pass 8} data, we search for new sources separately testing both a point or extended source at the location of the HAWC candidate. The spectrum of the source is modeled as a simple power law. After initially fitting a putative point source at the HAWC position, the position is re-localized. The normalization of other sources within 5$\degr$ are left as free parameters in the fit. To search for a possible extended source, we use a uniformly illuminated disk with a radius of 0.2$\degr$ as the initial spatial model. The $fermipy$ package \citep{2017arXiv170709551W} fits both the radius and centroid of the possible extended source. If a statistically significant source is not found, ULs at the position of the 2HWC source are computed at 95\% CL using a Bayesian method. The assumed spectral indices are 2.0, 3.0 and the index reported in the 2HWC catalog (see \autoref{HAWC_parameters}). Extended source ULs are computed assuming a radius of 0.16$\degr$, for comparison with the limits placed by MAGIC.

\begin{table*}
	\begin{center}
    	\caption{MAGIC differential ULs (at 95\% CL) for 2HWC J2006+341, 2HWC J1907+084$^{*}$ and 2HWC J1852+013$^{*}$ assuming a power-law spectrum with spectral index of  $\Gamma=2.64, 3.25$, and 2.90, respectively. ULs for both point-like ($\lesssim 0.10\degr$) and extended ($\sim 0.16\degr$ radius) assumptions are shown in each case. Due to low statistics, ULs at the highest energy ranges are not always computed for 2HWC J2006+341 and 2HWC J1907+084$^{*}$} 
    \label{table:differentialULs_MAGIC} 
	\begin{tabular}{ l c c | c c | c c}
		\hline
		\hline
		\multicolumn{1}{c}{Energy range} & \multicolumn{2}{c }{2HWC J2006+341} & \multicolumn{2}{c }{2HWC J1907+084$^{*}$} & \multicolumn{2}{c}{2HWC J1852+013$^{*}$}\T \\	
		\multicolumn{1}{c}{[GeV]} & \multicolumn{6}{c}{[photons cm$^{-2}$s$^{-1}$]} \T \\
	     \cline{2-7}
	     & Point-like & Extended & Point-like & Extended & Point-like & Extended\\
		\hline
			139.2 -- 300.0 & 2.6$\times10^{-11}$ & 6.2$\times10^{-11}$ & 7.1$\times10^{-11}$ & 3.1$\times10^{-10}$ & 1.7$\times10^{-11}$ & 4.6$\times10^{-11}$\\
			300.0 -- 646.3 & 1.4$\times10^{-12}$ & 1.0$\times10^{-11}$ & 8.0$\times10^{-12}$ & 1.7$\times10^{-11}$ & 8.6$\times10^{-13}$ & 4.9$\times10^{-12}$\\
			646.3 -- 1392.5 & 2.5$\times10^{-13}$ & 1.3$\times10^{-12}$ & 2.5$\times10^{-12}$ & 2.3$\times10^{-12}$ & 9.0$\times10^{-14}$ & 3.9$\times10^{-13}$ \\
			1392.5 -- 3000.0 & 6.0$\times10^{-14}$ & 9.9$\times10^{-14}$ & 1.7$\times10^{-13}$ & 1.0$\times10^{-12}$ & 6.7$\times10^{-14}$ & 1.2$\times10^{-13}$ \\
			3000.0 -- 6463.3 & 1.8$\times10^{-14}$ & 2.7$\times10^{-14}$ & -- & 1.4$\times10^{-13}$ & 7.6$\times10^{-15}$ & 1.3$\times10^{-14}$\\
			6463.3 -- 13924.8 & -- & 9.5$\times10^{-15}$ & -- & 1.3$\times10^{-14}$ & 1.1$\times10^{-14}$& 5.5$\times10^{-14}$\\ 
		\hline
	\end{tabular} 
	\end{center}
\end{table*}

\section{Observations and Results}
\label{section3}
\begin{figure*}
\begin{center}
		\includegraphics[width=1.\linewidth]{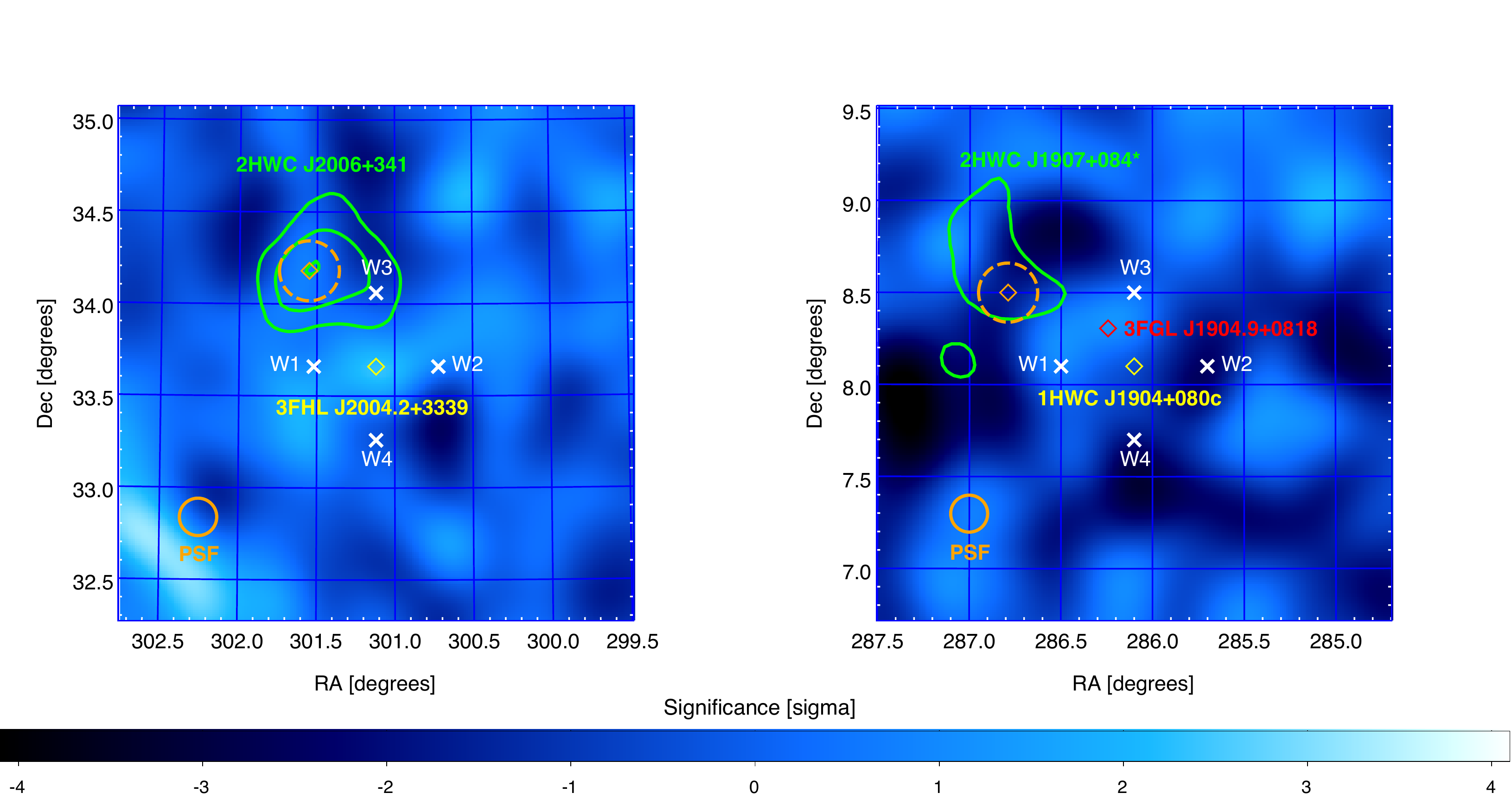}
		\caption{MAGIC significance skymaps for the FoV of 2HWC J2006+341 and 2HWC J1907+084$^{*}$, searching for sources of $\sim$ 0.16$\degr$ radius. The four different \textit{wobble} positions, to which MAGIC pointed during the 3FHL J2004.2+3339 and 1HWC J1904+080c observations, are tagged with W1, W2, W3 and W4 in white color. MAGIC PSF is shown in orange at the left bottom in each panel. \textit{Left panel:} Skymap of the observations at the direction of 3FHL J2004.2+3339 (yellow diamond). This FoV contains 2HWC J2006+341 (orange diamond) located at $\sim$ 0.63$\degr$ from the nominal position of MAGIC observations. Centered at the position of 2HWC J2006+341, the orange dashed circle corresponds to the assumed extension of 0.16$\degr$ used for the MAGIC analysis. The HAWC contours (4$\sigma$, 5$\sigma$ and 6$\sigma$) are shown as green solid lines. \textit{Right panel:} Skymap for the observations of the 1HWC J1904+080c (yellow diamond) FoV in which 2HWC J1907+084$^{*}$ (orange diamond) is enclosed. Dashed orange circle represents the 0.16$^{\circ}$ MAGIC extended assumption, while HAWC contours (at the level of $5\sigma$) are shown as green solid lines. The position of the closest \textit{Fermi}-LAT source, 3FGL J1904.9+0818, is marked as a red diamond.}
		\label{fig:skymap}
\end{center}
\end{figure*}

\begin{figure}
\begin{center}
		\includegraphics[width=1.\linewidth]{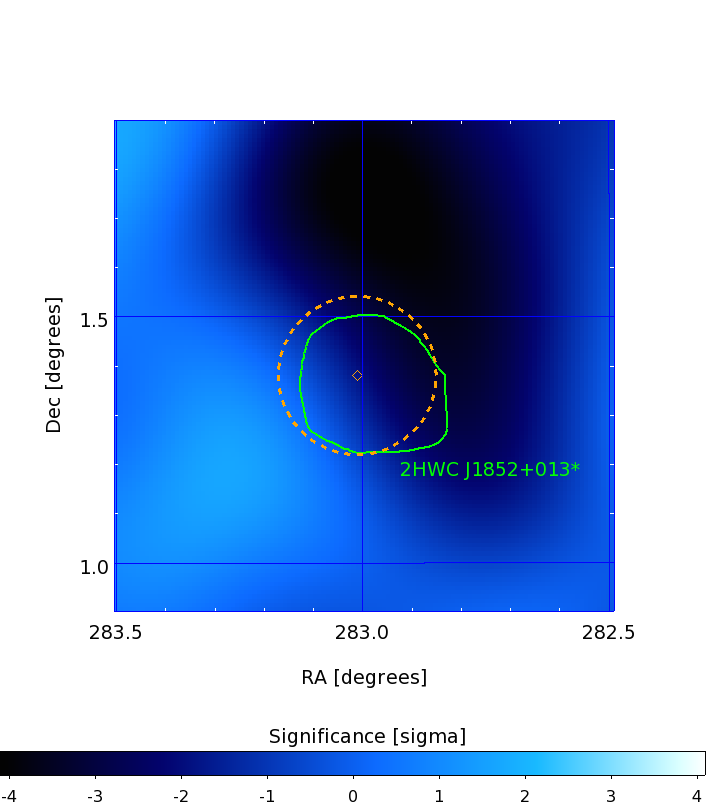}
		\caption{1$^{\circ} \times$ 1$^{\circ}$ MAGIC significance skymap, looking for 0.16$^{\circ}$ extended sources around 2HWC J1852+013$^{*}$, whose position is depicted as an orange diamond. Dashed orange circle shows the extension of 0.16$^{\circ}$ analyzed by MAGIC, whilst the green solid line corresponds to the 8$\sigma$ HAWC contour.}
		\label{fig:skymap_J1852}
\end{center}
\end{figure}

In the following section, we describe the different regions of the sky that contain the three selected sources, along with the corresponding observations and results. MAGIC differential ULs are listed in \autoref{table:differentialULs_MAGIC}, while \textit{Fermi}-LAT integral ULs (above 10 GeV) are quoted in \autoref{table:fermilatULs} for both point-like and extended hypotheses. \autoref{fig:skymap} presents the MAGIC significance skymaps for 2HWC J2006+341 and 2HWC J1907+084$^{*}$ assuming an extended source with a 0.16$^{\circ}$ radius. A smaller 1$^{\circ} \times$ 1$^{\circ}$ MAGIC significance skymap centered in 2HWC J1852+013$^{*}$ is shown in \autoref{fig:skymap_J1852}. The skymap of the entire FoV for this source will be included in a dedicated MAGIC paper on the surrounding region that is in preparation. The flat significance field displayed in all skymaps is compatible with background in the entire FoV. The multiwavelength spectral energy distribution (SED) for each 2HWC source is presented in \autoref{SED}. 

\subsection{2HWC J2006+341}
2HWC J2006+341 is in the FoV (at $\sim$ 0.63$^{\circ}$) of the compact radio/optical nebula G70.7+1.2, which is thought to be powered by a pulsar/binary system interacting with a surrounding molecular cloud. An unidentified source, 3FHL J2004.2+3339, was detected at the position of this putative binary system. Therefore, VHE gamma-ray emission from the G70.7+1.2 region could be expected due to the interaction between the pulsar wind with both the stellar wind of the companion star and the molecular cloud.

MAGIC observed 3FHL J2004.2+3339 with an extended range of zenith angles from 5$\degr$ to 50$\degr$. The total data sample amounts to $\sim 61$ hours of good quality data from April 2015 to August 2016. No significant signal is found in the direction of 2HWC J2006+341 as either a point-like or extended source. In order to calculate the integral flux ULs, MAGIC adopts a power-law distribution with photon index $\Gamma=2.64$, following the HAWC results. Under point-like assumption, the integral UL, computed at 95\% CL for energies greater than 300 GeV, is $4.0\times10^{-13}$ photons~cm$^{-2}$s$^{-1}$, while for $0.16\degr$ radius, it increases to $3.3\times10^{-12}$ photons~cm$^{-2}$s$^{-1}$. On the other hand, the integral UL for a point-like source in the direction of 3FHL J2004.2+3339 is $4.3\times10^{-13}$ photons~cm$^{-2}$s$^{-1}$ for energies above 300 GeV and assuming a power-law index of 2.6. In the GeV regime, no known \textit{Fermi} catalog source is found to be coincident with 2HWC J2006+341. The closest \textit{Fermi}-LAT source is the already mentioned 3FHL J2004.2+3339 (coincident within the errors with 3FGL J2004.4+338, \citealt{2015ApJS..218...23A}). This source has been searched extensively for pulsations \citep{2017ApJ...834..106C}. Using the analysis method described in Section \ref{Fermiinfo}, no significant (TS $\geq$ 25) source is detected using either the point or extended source models. 

\subsection{2HWC J1907+084$^{*}$}
2HWC J1907+084$^{*}$ is located at $\sim$ 0.79$\degr$ from 1HWC J1904+080c, included in the first HAWC survey. 1HWC J1904+080c was detected with a pre-trial significance of $5.14\sigma$, which motivated MAGIC follow-up observations. The coordinates were not coincident with any known TeV source, although it was close (at 0.3$\degr$) to a \textit{Fermi}-LAT gamma-ray hotspot ($<$5$\sigma$), 3FGL J1904.9+0818 \citep{2015ApJS..218...23A}. Nevertheless, its significance was only $\sim$ 4$\sigma$ after trials in the published 1HWC catalog \citep{2016ApJ...817....3A}. 

MAGIC performed observations  in the direction of 1HWC J1904+080c over 6 non-consecutive nights from May 10 2015 to May 19 2015. After rejecting the data affected by non-optimal weather conditions, the total amount of time reached in this FoV was 4.20 hours. The region was observed at medium zenith angles from 30$\degr$ to $50\degr$. No excess is found during the analysis of 2HWC J1907+084$^{*}$ data. The 95\% CL integral ULs for  $E>300$ GeV and index $\Gamma=3.25$ are $2.8\times10^{-12}$ photons~cm$^{-2}$s$^{-1}$ and $4.6\times10^{-12}$ photons~cm$^{-2}$s$^{-1}$ for the point-like and extended hypotheses, respectively. MAGIC does not find any significant excess at the position of 1HWC J1904+080c either, which leads to an integral flux UL for energies greater than 300 GeV of $4.1\times10^{-12}$ photons~cm$^{-2}$s$^{-1}$, assuming a power-law spectrum of $\Gamma=2.6$. From the \textit{Fermi}-LAT, the \texttt{Pass 8} analysis yields no significant emission in the direction of 2HWC J1907+084$^{*}$, either during the point-like or the extended analysis. 

\subsection{2HWC J1852+013$^{*}$}
2HWC J1852+013$^{*}$ is located in the FoV of the W44 SNR, as well as in the FoV of the established VHE sources HESS J1858+020 and HESS J1857+026 (subdivided in 2 emission sites MAGIC J1857.2+0263 and MAGIC J1857.6+0297; \citealt{2014A&A...571A..96M}). The region was thus extensively observed by the MAGIC collaboration. 2HWC J1852+013$^{*}$ is also located at 0.56$\degr$ away from 3FGL J1852.8+0158, which is classified as a probable young pulsar using machine learning techniques \citep{2016ApJ...820....8S}.

The dataset used by MAGIC here comprises approximately $120$ hours of dark quality data, taken from April 2013 to June 2014, with a span in zenith range from  $25\degr$ to $50\degr$. MAGIC does not find any excess in the direction of 2HWC J1852+013$^{*}$. Adopting $\Gamma=2.90$, the constraining 95\% CL integral ULs are $3.8\times10^{-13}$ photons cm$^{-2}$s$^{-1}$ for the point-like search and $1.7\times10^{-12}$ photons cm$^{-2}$s$^{-1}$ for the extended search. Specific background selection using OfWP was applied in this case, ensuring that no background control region overlaps with any of the several VHE emitting sources in the FoV. As per the previous sources, ULs are given for $E > 300$ GeV. Neither cataloged nor new sources from the \texttt{Pass 8} analysis arises in the \textit{Fermi}-LAT analysis of 2HWC J1852+013$^{*}$. 

\begin{table*}
\centering
\caption{\textit{Fermi}-LAT 95\% CL flux ULs, above 10 GeV, assuming point-like source and extended source a radius of 0.16$^{\circ}$.} 
\hfill{}
\label{table:fermilatULs}
	\begin{tabular}{ l c c | c c | c c }
    	\cline{2-7}
        & \multicolumn{2}{c |}{$\Gamma_{\rm HAWC}$} & \multicolumn{2}{ c |}{$\Gamma$=2.0} & \multicolumn{2}{c}{$\Gamma$ = 3.0} \T \\
        \cline{2-7}
	    & Point-like & Extended & Point-like & Extended & Point-like & Extended \\
		& \multicolumn{2}{c |}{[$\times10^{-11}$ photons cm$^{-2}$s$^{-1}$]} & \multicolumn{2}{  c | }{[$\times10^{-11}$ photons cm$^{-2}$s$^{-1}$]} & \multicolumn{2}{c}{[$\times10^{-11}$ photons cm$^{-2}$s$^{-1}$]}\T \T \\	     
		\hline
        J2006+341 & 2.4 & 4.4 & 2.3 & 4.7 & 2.4 & 4.2 \\
        J1907+084* & 3.1 & 3.1 & 2.7 & 2.7 & 3.2 & 3.2 \\
        J1852+013* & 2.1 & 3.7 & 2.0 & 3.3 & 2.0 & 3.7 \\
		\hline
	\end{tabular}
	\hfill{}
\end{table*}

\section{Discussion}
\label{section4}
Given that the largest population of TeV emitters in our Galaxy are pulsar wind nebulae (PWNe; see e.g. \citealt{HGPS_paper}), the selected candidates may be expected to be this source type. However, the lack of a detection by either MAGIC or \textit{Fermi}-LAT complicates the identification of these sources. In order to investigate their possible PWN nature, we look for detected pulsars near these 2HWC sources using the ATNF catalog\footnote{http://www.atnf.csiro.au/people/pulsar/psrcat/} \citep{ATNFcatalog}. According to the characteristic ages of the pulsars around the three selected 2HWC sources (all above a few tens of kyr), if these pulsars had high initial kick velocities, they could now be significantly offset from their initial positions and have left behind an old PWN with no compact object powering it. In this case, the pulsar position is shifted from the PWN, and without injection of magnetic flux, the nebula's emission is expected to be dominated by inverse Compton(IC). 

PSR J2004+3429 is the closest known pulsar to  2HWC J2006+341 at a separation of $0.40\degr$, and is the only one within a 1$\degr$ radius. This pulsar lies at a distance of 10 kpc, displays a spin-down power of $\dot{E}=5.8\times10^{35}$ erg s$^{-1}$ and has a characteristic age of $\tau=18$ kyr. Although energetic enough to power a TeV PWN (see \citealt{HESSPWNePop}), the distance between 2HWC J2006+341 and PSR J2004+3429 makes this connection improbable: given the characteristic age of 18 kyr, an offset of 0.40$\degr$ ($\sim 70$ pc) could only be explained with an improbably large kick velocity for the pulsar of $\sim 4000$ km~s$^{-1}$. The mean 2-D speed for both young and old ($< 3$ Myr) pulsars was determined to be only $307\pm47$ km~s$^{-1}$ by \cite{Hobbs2005} with a study involving a subsample of $\sim 50$ pulsars' proper motion. The offset may be cosiderably less when considering HAWC systematic and statistical errors on the 2HWC source location of 0.40$\degr \pm$~0.10$\degr _{syst} \pm$~0.13$\degr _{stat}$. Assuming the most constraining possible value, 0.24$\degr$, the necessary kick velocity would decrease  to $\sim 2300$ km~s$^{-1}$. This value is not far away from the fastest known pulsar at $\sim 1500$ km~s$^{-1}$ \citep{Hobbs2005}, though that value is also uncertain given the distance model applied. The highest speed for a pulsar with a well-measured distance is only 640 km~s$^{-1}$. Therefore, we conclude that it is unlikely that PSR J2004+3429 is directly responsible for the emission detected by HAWC. On the other hand, \cite{Linden2017} evaluated the probability of random association between 15 2HWC sources and their nearby pulsars, including 2HWC J2006+341 and PSR J2004+3429. For this case, they obtained a chance overlap of only 8\% (assuming a source extension of 0.9$^{\circ}$ as provided in the 2HWC catalog by assigning the halo-like structures visible in the residual skymaps to 2HWC J2006+341, which presents its own uncertainties). 

There are two pulsars within $0.50\degr$ of 2HWC J1907+084$^{*}$: PSR J1908+0833 at 0.30$\degr$, and PSR J1908+0839 at $0.33\degr$. The former is located at a distance of $\sim 11$ kpc, with a characteristic age of $\tau=4.1$ Myr and a spin-down power of $\dot{E}=5.8\times10^{32}$ erg s$^{-1}$. The very low spin-down power and old age make it unlikely to be currently powering a TeV PWN. Alternatively, PSR J1908+0839, at 8.3 kpc and with a characteristic age of $\tau=1.2$ Myr, is more energetic with $\dot{E}=1.5\times10^{34}$ erg s$^{-1}$, so a relation between this pulsar and the 2HWC source cannot be initially ruled out.  As done for 2HWC J2006+341, we calculate the kick velocity for the pulsar, now with an offset of 0.33$\degr$ and a characteristic age of 1.2 Myr. The obtained velocity is $\sim 40$ km~s$^{-1}$, which is low relative to the average kick velocity observed through proper motion studies but remains to be a valid possibility (see Figure 4b from \citealt{Hobbs2005}). This velocity stays within the young pulsars' 2-D speed distribution even considering HAWC uncertanties (0.33$\degr \pm$~0.10$\degr _{syst} \pm$~0.27$\degr _{stat}$). We can compare PSR J1908+0839 with the pulsar hosted by Geminga, a well-known TeV PWN detected by Milagro \citep{MilagroGeminga} and recently by HAWC \citep{Geminga_HAWC}. Geminga's pulsar displays a spin-down power of $\dot{E}=3.25\times10^{34}$ erg s$^{-1}$, very similar to that shown by PSR J1908+0839, but its distance is $\sim30$ times smaller ($d_{Geminga}=250$ pc). Given the similar spin-down power, the nebula PSR J1908+0839 powers should also have a comparable luminosity with respect to Geminga's PWN, which would lead to a flux around three orders of magnitude smaller than the flux of Geminga and undetectable by HAWC. Therefore, it is unlikely that 2HWC J1907+084$^{*}$ and PSR J1908+0839 are associated with one another.

Finally, the closest pulsar to the source 2HWC J1852+013$^{*}$ is PSR J1851+0118, offset by only $0.10\degr$. This pulsar lies at a distance of 5.6 kpc and has a characteristic age of $\sim100$ kyr \citep{2017ApJ...835...29Y}. If both objects are related the required pulsar velocity is a reasonable value of $\sim 100$ km~s$^{-1}$, though this may be as high as $\sim 245$ km~s$^{-1}$ when considering the largest offset given by 0.10$\degr \pm$~0.10$\degr _{syst} \pm$~0.13$\degr _{stat}$. However, this pulsar has a relatively low spin-down power, $\dot{E}=7.2\times10^{33}$ erg s$^{-1}$, which along with its high characteristic age make it unlikely to accelerate particles that can emit gamma rays in the TeV regime. To quantitatively test this scenario, we use the \texttt{Naima} software\footnote{\url{naima.readthedocs.org}} to model the relativistic parent population of the non-thermal gamma-ray emission accounting for different radiative models (see \citealt{Naima}). To obtain the emissivity of the electron population that gave rise to the gamma-ray emission seen by HAWC, we assume that IC is the dominant radiative process and that the electron spectrum is defined by a simple power law. The target photon field for this process is expected to be a combination of cosmic microwave background (CMB) and infrared (IR) photons. The assumed energy densities in each case are standard Galactic values of $u_{CMB}=0.25$ eV cm$^{-3}$ and $u_{IR}=0.30$ eV cm$^{-3}$. With such features, the total energy carried by electrons above $\sim 10$ TeV needed to explain HAWC detection would be $W_{e} (> 10~\textrm{TeV})\sim 6.0 \times 10^{46}$ erg. 

Alternatively, we consider the cooling time of these electrons ($t_{cool}$), which is computed as follows \citep{AharonianBook}:
\begin{equation}
\label{eq:coolingtimeelectrons}
t_{cool} = 3 \cdot 10^{8} \left(\frac{E_{e}}{\textrm{GeV}}\right)^{-1} \left(\frac{u}{\textrm{eV/cm}^{3}}\right)^{-1}~[\textrm{yr}]
\end{equation}
where $E_{e}=10$ TeV is the electron energy and $u$ is the total energy density of the medium. In this case, we account for IC and synchrotron losses and hence, assuming a temperature of $\sim$ 25 K for the IR photon field \citep{Moderski2005}, $u$ can be described as:
\begin{equation}
u \simeq \frac{B^{2}}{8\pi} + u_{CMB}\left(1+0.01\cdot \frac{E_{e}}{\textrm{TeV}}\right)^{-3/2} + u_{IR}\left(1+0.1\cdot \frac{E_{e}}{\textrm{TeV}}\right)^{-3/2}
\end{equation}
where $B$ is the magnetic field. We use the aforementioned energy densities for the CMB and IR photon fields, and for the magnetic field we assume the minimum possible value given by the interstellar magnetic field, $B=3~\mu$G, as there is no measured value for the source. Under these assumptions, the cooling time is $t_{cool}\sim 57$ kyr. Given the spin-down power of PSR J1851+0118, $\dot{E}=7.2\times10^{33}$ erg s$^{-1}$, the total energy released by the pulsar during the $t_{cool}$ period would be $W'_{e}\sim 1.3\times 10^{46}$ erg. Consequently, even assuming that all the energy released by the pulsar was used to accelerate electrons above 10 TeV, there would not be enough energy to power a PWN with the gamma-ray brightness detected by HAWC. According to our model, such a PWN would require an energy injection greater than $\sim 6.0 \times 10^{46}$ erg, which is already higher than $W'_{e}$. The low $B$ and $u_{IR}$ used in this calculation of $t_{cool}$ provide maximum values for both $t_{cool}$ and the injected pulsar energy, $W'_{e}$. Higher $u_{IR}$ value would produce higher losses and therefore, a smaller $W_{e}$. However, an extremely high value for $u_{IR}$ (well above $2$ eV cm$^{-3}$, the IR energy density observed around Cassiopea A and one of the highest for a Galactic TeV source) would be needed to decrease $W_{e}$ below $10^{46}$ erg. We do not consider more complex scenarios in which $\dot{E}$ or $B$ change with time. The same parent population study was applied to 2HWC J1907+084$^{*}$ and we reach the same conclusions that corroborated the non-relation with the surrounding pulsars. 

\begin{figure}
\centering
   \begin{subfigure}{0.5\linewidth} \centering
   \adjincludegraphics[trim={0 0 0 0},clip,width=\columnwidth]{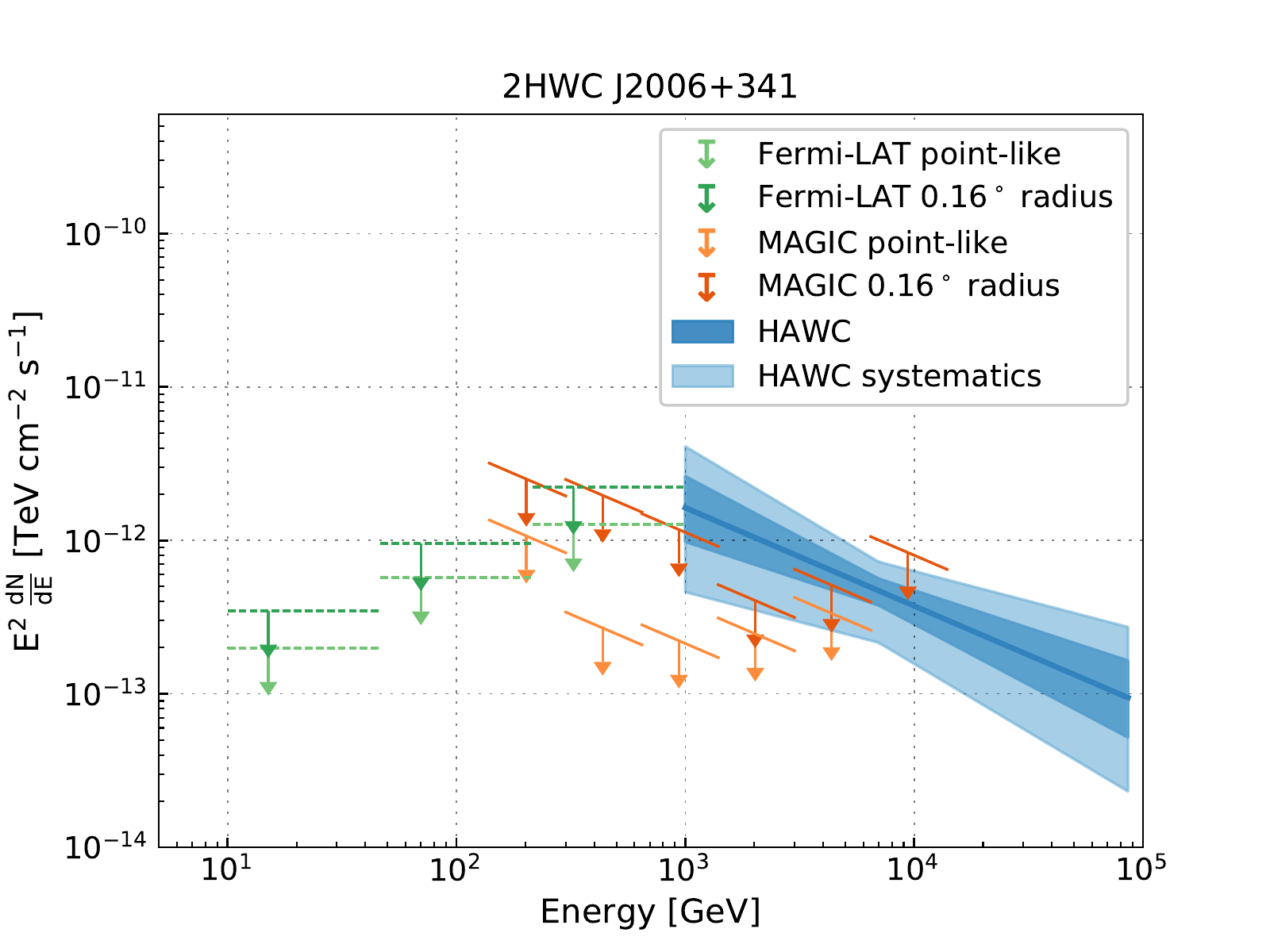}
   \label{fig:SEDJ2006}
   \vspace{0.5 cm}   
      \adjincludegraphics[trim={0 0 0 0},clip,width=\columnwidth]{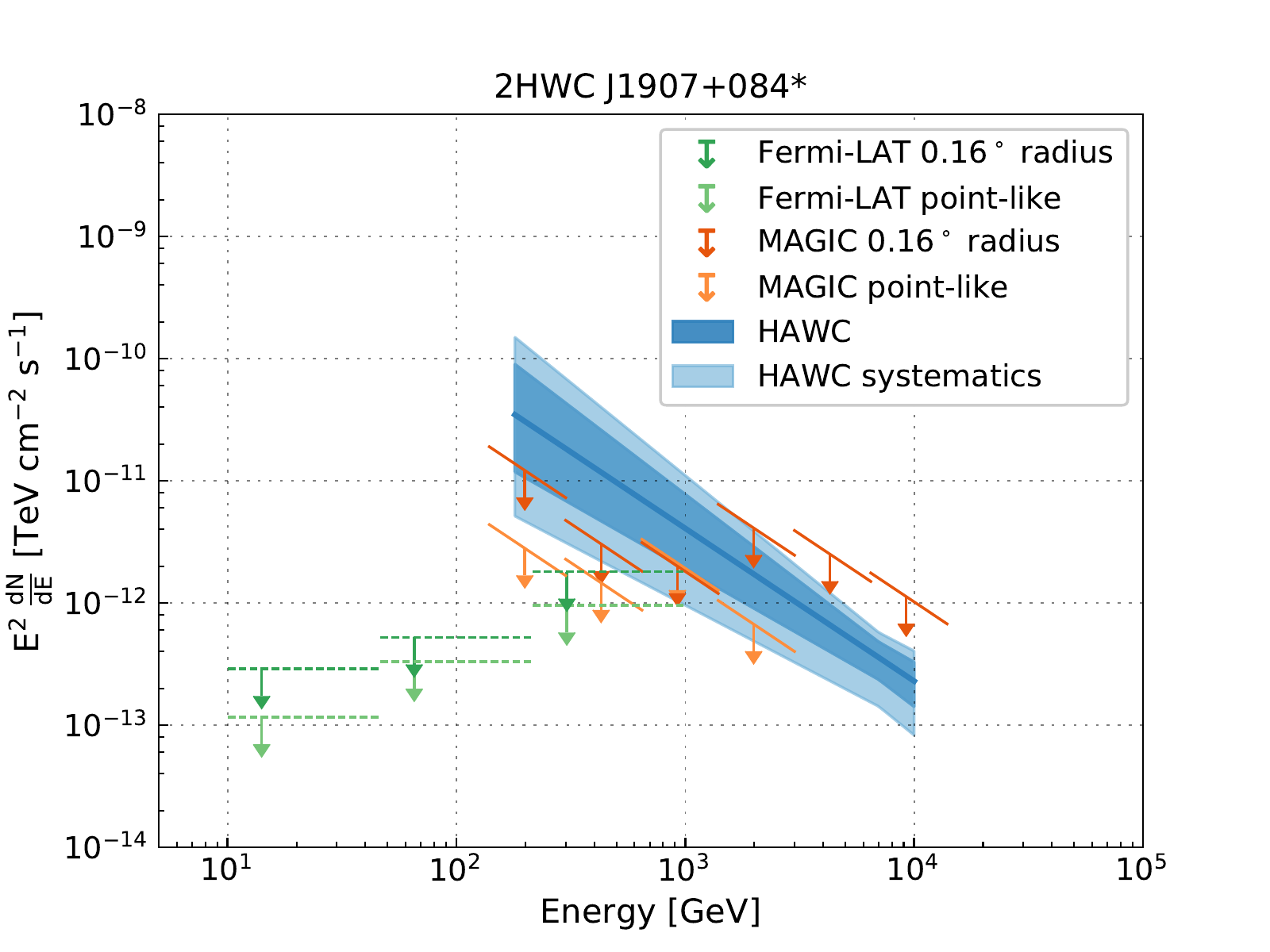}
   \label{fig:SEDJ1907}
   \end{subfigure}
   \begin{subfigure}{0.5\linewidth} \centering
         \adjincludegraphics[trim={0 0 0 0},clip,width=\columnwidth]{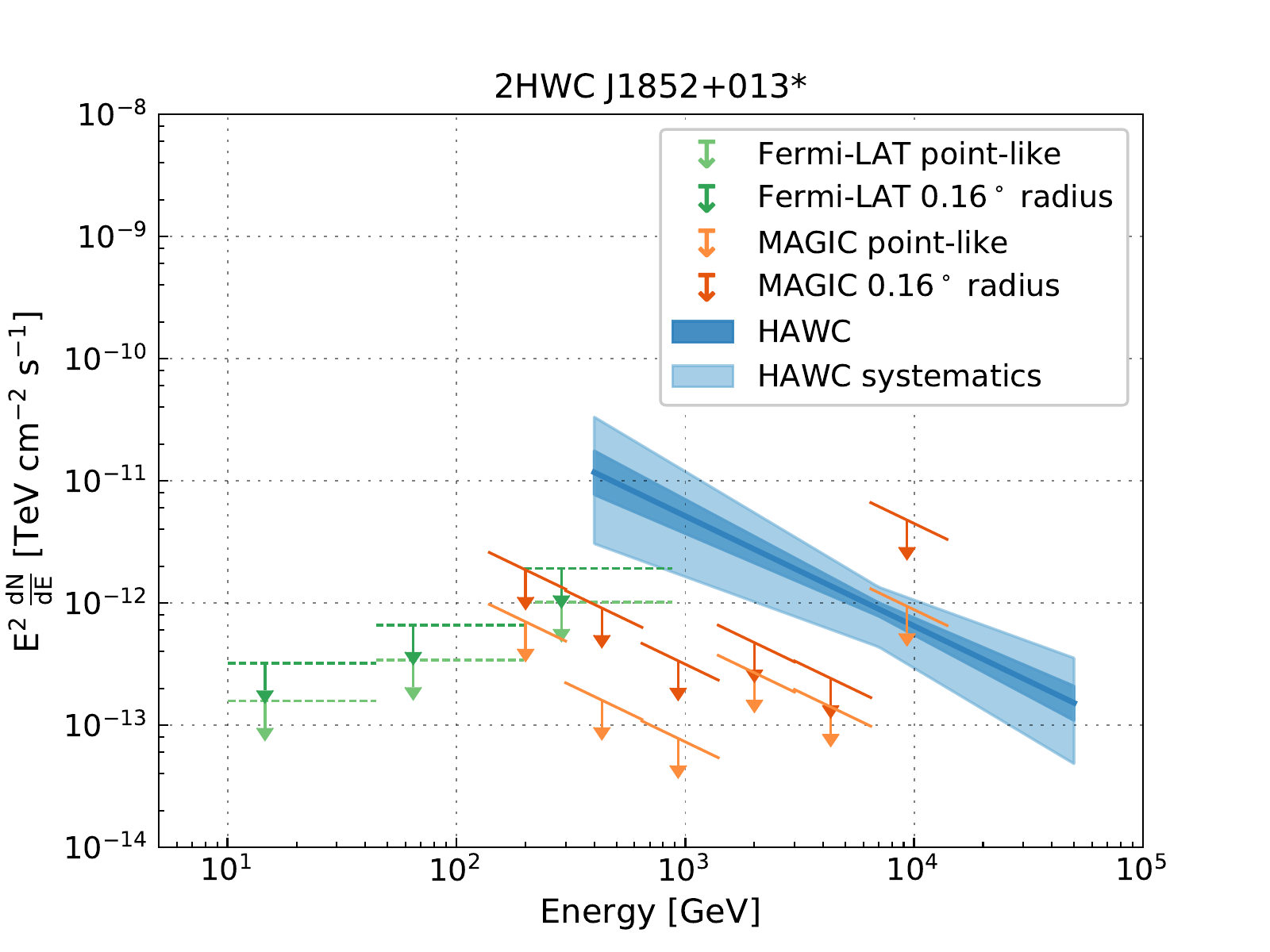}
   \label{fig:SEDJ1852}
   \end{subfigure}
   \caption{Spectral energy distribution from 10 GeV up to $\sim 90$ TeV. In all cases, the assumed spectrum for the sources is a power-law function with photon index $\Gamma=2.64$ for 2HWC J2006+341 (\textit{top}), $\Gamma=3.25$ for 2HWC J1907+084$^{*}$ (\textit{middle}) and $\Gamma=2.90$ for 2HWC J1852+013$^{*}$ (\textit{bottom}), as obtained by HAWC (see \autoref{HAWC_parameters}). \textit{Fermi}-LAT 95\% confidence level ULs for $0.16\degr$ disk and point-like hypotheses are shown with dashed green and light green lines, respectively. MAGIC 95\% confidence level ULs are displayed for both a point-like hypothesis (light orange) and a $0.16\degr$ radial extension (orange). The HAWC spectrum (dark blue) is obtained for the parameters given in \autoref{HAWC_parameters}. The light blue band indicates the HAWC spectrum taking into account 1$\sigma$ systematic errors of 0.2 and 50\% in the photon index and flux, respectively.} 
\label{SED}
\end{figure}

MAGIC and LAT ULs also help to constrain our understanding of the spectrum and morphology of these HAWC sources. 
The SEDs for the three candidate PWNe are shown in \autoref{SED}. MAGIC and \textit{Fermi}-LAT analyses are computed with the photon index provided by HAWC (see \autoref{HAWC_parameters}). In the cases of 2HWC J2006+341 and 2HWC J1907+084$^{*}$, the MAGIC {\bf and LAT} extended ULs are at the level of the HAWC spectrum considering HAWC systematic errors of 0.2 in the photon index and 50\% in the flux normalization. However, point-like hypotheses are in contradiction with HAWC results below energies of $\sim 4$ TeV and $\sim 900$ GeV, respectively. Therefore, it is expected that these two 2HWC sources are extended, with at least a radius of $\sim 0.16\degr$. On the other hand, both MAGIC and \textit{Fermi}-LAT results on 2HWC J1852+013$^{*}$ are incompatible with the HAWC spectrum below energies of $\sim 10$ TeV.

These results can be understood in two ways: 2HWC J1852+013$^{*}$ is much more extended than the assumed radius of $0.16\degr$, which would increase MAGIC and \textit{Fermi}-LAT ULs above the flux estimated by HAWC; or the source does not emit in the sub-TeV regime, consistent with the constraining ULs obtained by both MAGIC and the LAT. In the later case, the spectral shaped of 2HWC J1852+013$^{*}$ would have a harder spectrum in the sub-TeV regime, and a minimum energy of around 10 TeV, instead of 400 GeV, should be assumed (see \autoref{HAWC_parameters}). To constrain the former case, we calculated LAT ULs for disks of larger radii. For a disk of $1.0\degr$ radius the LAT UL at energies $> 0.2$ TeV is within 1$\sigma$ statistical errors of the measured HAWC flux, extrapolated to lower energies. However, this would also require a harder spectrum in the GeV regime so as to not exceed LAT ULs at lower energies. Additionally, as reported in the 2HWC catalog \cite{2HWC_Catalog}, there may be a significant contribution from diffuse galactic emission at the location of 2HWC J1852+013* to which HAWC would be sensitive and MAGIC would not.

\section{Conclusion}
\label{sec:conclusions}
After the release of the 2HWC catalog, MAGIC and \textit{Fermi}-LAT performed dedicated analyses on three new TeV sources detected by the wide FoV observatory HAWC. None of them were detected at lower energies and no hotspot was found near them. However, owing to the increased time and good quality data of most of the MAGIC and the \textit{Fermi}-LAT observations, constraints on the extension of the sources were possible. With this aim, we performed both point-like and extended source searches. For 2HWC J2006+341 and 2HWC J1907+084$^{*}$, a radius of $\sim 0.16\degr$ is viable given limits from the extended source search by MAGIC. For 2HWC J1852+013$^{*}$, MAGIC and \textit{Fermi}-LAT results with respect to HAWC spectra suggest a much larger extension or a harder spectrum below $\sim 10$ TeV. Moreover, we find that none of the known pulsars in the vicinity of 2HWC J2006+341, 2HWC J1907+084$^{*}$ or 2HWC J1852+013$^{*}$ are likely to directly power these objects. It may be that these 2HWC sources are PWN created by as yet un-detected pulsars, or have some other origin such as a Galactic supernova remnant. More sensitive observations in the near future will provide valuable information on the nature of these sources and help to disentangle features in the crowded regions. 

\section*{Acknowledgements}
The MAGIC Collaboration would like to thank the Instituto de Astrof\'{\i}sica de Canarias for the excellent working conditions at the Observatorio del Roque de los Muchachos in La Palma. The financial support of the German BMBF and MPG, the Italian INFN and INAF, the Swiss National Fund SNF, the ERDF under the Spanish MINECO (FPA2015-69818-P, FPA2012-36668, FPA2015-68378-P, FPA2015-69210-C6-2-R, FPA2015-69210-C6-4-R, FPA2015-69210-C6-6-R, AYA2015-71042-P, AYA2016-76012-C3-1-P, ESP2015-71662-C2-2-P, CSD2009-00064), and the Japanese JSPS and MEXT is gratefully acknowledged. This work was also supported by the Spanish Centro de Excelencia ``Severo Ochoa'' SEV-2012-0234 and SEV-2015-0548, and Unidad de Excelencia ``Mar\'{\i}a de Maeztu'' MDM-2014-0369, by the Croatian Science Foundation (HrZZ) Project 09/176 and the University of Rijeka Project 13.12.1.3.02, by the DFG Collaborative Research Centers SFB823/C4 and SFB876/C3, and by the Polish MNiSzW grant 745/N-HESS-MAGIC/2010/0.\\
The HAWC Collaboration acknowledges the support from: the US National Science Foundation (NSF) the US Department of Energy Office of High-Energy Physics; the Laboratory Directed Research and Development (LDRD) program of Los Alamos National Laboratory; Consejo Nacional de Ciencia y Tecnolog\'{\i}a (CONACyT), M{\'e}xico (grants 271051, 232656, 260378, 179588, 239762, 254964, 271737, 258865, 243290, 132197, 281653)(C{\'a}tedras 873, 1563), Laboratorio Nacional HAWC de rayos gamma;
L'OREAL Fellowship for Women in Science 2014;
Red HAWC, M{\'e}xico; DGAPA-UNAM (grants IG100317, IN111315, IN111716-3, IA102715, 109916, IA102917, IN112218); VIEP-BUAP;
PIFI 2012, 2013, PROFOCIE 2014, 2015; the University of Wisconsin Alumni Research Foundation; the Institute of Geophysics, Planetary Physics, and Signatures at Los Alamos National Laboratory; Polish Science Centre grant DEC-2014/13/B/ST9/945; Coordinaci{\'o}n de la Investigaci{\'o}n Cient\'{\i}fica de la Universidad Michoacana; Royal Society - Newton Advanced Fellowship 180385. Thanks to Scott Delay, Luciano D\'{\i}az and Eduardo Murrieta for technical support.\\
The \textit{Fermi}-LAT Collaboration acknowledges generous ongoing support from a number of agencies and institutes that have supported both the development and the operation of the LAT as well as scientific data analysis. These include the National Aeronautics and Space Administration and the Department of Energy in the United States, the Commissariat \`a l'Energie Atomique and the Centre National de la Recherche Scientifique / Institut National de Physique Nucl\'eaire et de Physique des Particules in France, the Agenzia Spaziale Italiana and the Istituto Nazionale di Fisica Nucleare in Italy, the Ministry of Education, Culture, Sports, Science and Technology (MEXT), High Energy Accelerator Research Organization (KEK) and Japan Aerospace Exploration Agency (JAXA) in Japan, and the K.~A.~Wallenberg Foundation, the Swedish Research Council and the Swedish National Space Board in Sweden. This work performed in part under DOE Contract DE-AC02-76SF00515.\\



\bibliographystyle{mnras}
\bibliography{bibliography_MFH} 

\onecolumn 

\twocolumn

\vspace*{0.5cm}
\noindent
$^{1}$ {ETH Zurich, CH-8093 Zurich, Switzerland} \\
$^{2}$ {Universit\`a di Udine, and INFN Trieste, I-33100 Udine, Italy} \\
$^{3}$ {National Institute for Astrophysics (INAF), I-00136 Rome, Italy} \\
$^{4}$ {Universit\`a di Padova and INFN, I-35131 Padova, Italy} \\
$^{5}$ {Technische Universit\"at Dortmund, D-44221 Dortmund, Germany} \\
$^{6}$ {Croatian MAGIC Consortium: University of Rijeka, 51000 Rijeka, University of Split - FESB, 21000 Split,  University of Zagreb - FER, 10000 Zagreb, University of Osijek, 31000 Osijek and Rudjer Boskovic Institute, 10000 Zagreb, Croatia.} \\
$^{7}$ {Saha Institute of Nuclear Physics, HBNI, 1/AF Bidhannagar, Salt Lake, Sector-1, Kolkata 700064, India} \\
$^{8}$ {Max-Planck-Institut f\"ur Physik, D-80805 M\"unchen, Germany} \\
$^{9}$ {Now at Centro Brasileiro de Pesquisas F\'isicas (CBPF), 22290-180 URCA, Rio de Janeiro (RJ), Brasil} \\
$^{10}$ {Unidad de Part\'iculas y Cosmolog\'ia (UPARCOS), Universidad Complutense, E-28040 Madrid, Spain} \\
$^{11}$ {Inst. de Astrof\'isica de Canarias, E-38200 La Laguna, and Universidad de La Laguna, Dpto. Astrof\'isica, E-38206 La Laguna, Tenerife, Spain} \\
$^{12}$ {University of \L\'od\'z, Department of Astrophysics, PL-90236 \L\'od\'z, Poland} \\
$^{13}$ {Deutsches Elektronen-Synchrotron (DESY), D-15738 Zeuthen, Germany} \\
$^{14}$ {Institut de F\'isica d'Altes Energies (IFAE), The Barcelona Institute of Science and Technology (BIST), E-08193 Bellaterra (Barcelona), Spain} \\
$^{15}$ {Universit\`a  di Siena and INFN Pisa, I-53100 Siena, Italy} \\
$^{16}$ {Universit\"at W\"urzburg, D-97074 W\"urzburg, Germany} \\
$^{17}$ {Finnish MAGIC Consortium: Tuorla Observatory and Finnish Centre of Astronomy with ESO (FINCA), University of Turku, Vaisalantie 20, FI-21500 Piikki\"o, Astronomy Division, University of Oulu, FIN-90014 University of Oulu, Finland} \\
$^{18}$ {Departament de F\'isica, and CERES-IEEC, Universitat Aut\'onoma de Barcelona, E-08193 Bellaterra, Spain} \\
$^{19}$ {Universitat de Barcelona, ICC, IEEC-UB, E-08028 Barcelona, Spain} \\
$^{20}$ {Japanese MAGIC Consortium: ICRR, The University of Tokyo, 277-8582 Chiba, Japan; Department of Physics, Kyoto University, 606-8502 Kyoto, Japan; Tokai University, 259-1292 Kanagawa, Japan; The University of Tokushima, 770-8502 Tokushima, Japan} \\
$^{21}$ {Inst. for Nucl. Research and Nucl. Energy, Bulgarian Academy of Sciences, BG-1784 Sofia, Bulgaria} \\
$^{22}$ {Universit\`a di Pisa, and INFN Pisa, I-56126 Pisa, Italy} \\
$^{23}$ Humboldt University of Berlin, Institut f\"ur Physik D-12489 Berlin Germany \\
$^{24}$ also at Dipartimento di Fisica, Universit\`a di Trieste, I-34127 Trieste, Italy \\
$^{25}$ also at Port d'Informaci\'o Cient\'ifica (PIC) E-08193, Bellaterra (Barcelona) Spain \\
$^{26}${Physics Division, Los Alamos National Laboratory, Los Alamos, NM, USA }\\
$^{27}${Instituto de F\'{i}sica, Universidad Nacional Aut\'{o}noma de M\'{e}xico, Ciudad de M\'{e}xico, Mexico }\\
$^{28}${Universidad Aut\'{o}noma de Chiapas, Tuxtla Guti\'{e}rrez, Chiapas, Mexico}\\
$^{29}${Universidad Michoacana de San Nicol\'{a}s de Hidalgo, Morelia, Mexico }\\
$^{30}${Department of Physics, Pennsylvania State University, University Park, PA, USA }\\
$^{31}${Department of Physics \& Astronomy, University of Rochester, Rochester, NY , USA }\\
$^{32}${Instituto de Astronom\'{i}a, Universidad Nacional Aut\'{o}noma de M\'{e}xico, Ciudad de M\'{e}xico, Mexico }\\
$^{33}${Department of Physics, University of Wisconsin-Madison, Madison, WI, USA }\\
$^{34}${Instituto Nacional de Astrof\'{i}sica, \'{O}ptica y Electr\'{o}nica, Puebla, Mexico }\\
$^{35}${Institute of Nuclear Physics Polish Academy of Sciences, PL-31342 IFJ-PAN, Krakow, Poland }\\
$^{36}${Facultad de Ciencias F\'{i}sico Matem\'{a}ticas, Benem\'{e}rita Universidad Aut\'{o}noma de Puebla, Puebla, Mexico }\\
$^{37}${Departamento de F\'{i}sica, Centro Universitario de Ciencias Exactas e Ingenier\'{i}as, Universidad de Guadalajara, Guadalajara, Mexico }\\
$^{38}${School of Physics, Astronomy, and Computational Sciences, George Mason University, Fairfax, VA, USA }\\
$^{39}${Department of Physics, University of Maryland, College Park, MD, USA }\\
$^{40}${Instituto de Geof\'{i}sica, Universidad Nacional Aut\'{o}noma de M\'{e}xico, Ciudad de M\'{e}xico, Mexico }\\
$^{41}${Department of Physics, Michigan Technological University, Houghton, MI, USA }\\
$^{42}${NASA Marshall Space Flight Center, Astrophysics Office, Huntsville, AL 35812, USA}\\
$^{43}${Max-Planck Institute for Nuclear Physics, 69117 Heidelberg, Germany}\\
$^{44}${Dept of Physics and Astronomy, University of New Mexico, Albuquerque, NM, USA }\\
$^{45}${School of Physics and Center for Relativistic Astrophysics - Georgia Institute of Technology, Atlanta, GA, USA 30332 }\\
$^{46}${Department of Physics and Astronomy, Michigan State University, East Lansing, MI, USA }\\
$^{47}${Universidad Politecnica de Pachuca, Pachuca, Hgo, Mexico }\\
$^{48}${Centro de Investigaci\'{o}n en Computaci\'{o}n, Instituto Polit\'{e}cnico Nacional, Mexico City, Mexico}\\
$^{49}${Physics Department, Centro de Investigaci\'{o}n y de Estudios Avanzados del IPN, Mexico City, DF, Mexico }\\
$^{50}${Universidad Aut\'{o}noma del Estado de Hidalgo, Pachuca, Mexico }\\
$^{51}${Instituto de Ciencias Nucleares, Universidad Nacional Aut\'{o}noma de M\'{e}xico, Ciudad de M\'{e}xico, Mexico }\\
$^{52}${Department of Physics and Astronomy, University of Utah, Salt Lake City, UT, USA }\\
$^{53}${Santa Cruz Institute for Particle Physics, University of California, Santa Cruz, Santa Cruz, CA, USA }\\
$^{54}${Department of Physics, Stanford University: Stanford, CA 94305–4060, USA}\\
$^{55}${Department of Physics and Astronomy. University of California. Irvine, CA 92697, USA}\\
$^{56}${Department of Physics and Astronomy, Clemson University, Kinard Lab of Physics, Clemson, SC 29634-0978, USA}\\ 
$^{57}${Universit\`a di Pisa and Istituto Nazionale di Fisica Nucleare, Sezione di Pisa I-56127 Pisa, Italy}\\
$^{58}${Istituto Nazionale di Fisica Nucleare, Sezione di Trieste, I-34127 Trieste, Italy}\\
$^{59}${Dipartimento di Fisica, Universit\`a di Trieste, I-34127 Trieste, Italy}\\
$^{60}${California State University, Los Angeles, Department of Physics and Astronomy, Los Angeles, CA 90032, USA}\\
$^{61}${Dipartimento di Fisica ``M. Merlin" dell'Universit\`a e del Politecnico di Bari, I-70126 Bari, Italy}\\
$^{62}${Istituto Nazionale di Fisica Nucleare, Sezione di Bari, I-70126 Bari, Italy}\\
$^{63}${W. W. Hansen Experimental Physics Laboratory, Kavli Institute for Particle Astrophysics and Cosmology, Department of Physics and SLAC National Accelerator Laboratory, Stanford University, Stanford, CA 94305, USA}\\
$^{64}${Istituto Nazionale di Fisica Nucleare, Sezione di Torino, I-10125 Torino, Italy}\\
$^{65}${Dipartimento di Fisica, Universit\`a degli Studi di Torino, I-10125 Torino, Italy}\\
$^{66}${Department of Physics and Astronomy, University of Padova, Vicolo Osservatorio 3, I-35122 Padova, Italy}\\
$^{67}${NASA Goddard Space Flight Center, Greenbelt, MD 20771, USA}\\
$^{68}${Laboratoire Univers et Particules de Montpellier, Universit\'e Montpellier, CNRS/IN2P3, F-34095 Montpellier, France}\\
$^{69}${Laboratoire Leprince-Ringuet, \'Ecole polytechnique, CNRS/IN2P3, F-91128 Palaiseau, France}\\
$^{70}${Center for Research and Exploration in Space Science and Technology (CRESST) and NASA Goddard Space Flight Center, Greenbelt, MD 20771, USA}\\
$^{71}${INAF-Istituto di Astrofisica Spaziale e Fisica Cosmica Milano, via E. Bassini 15, I-20133 Milano, Italy}\\
$^{72}${Harvard-Smithsonian Center for Astrophysics, Cambridge, MA 02138, USA}\\
$^{73}${Italian Space Agency, Via del Politecnico snc, 00133 Roma, Italy}\\
$^{74}${Space Science Data Center - Agenzia Spaziale Italiana, Via del Politecnico, snc, I-00133, Roma, Italy}\\
$^{75}${Istituto Nazionale di Fisica Nucleare, Sezione di Perugia, I-06123 Perugia, Italy}\\
$^{76}${Dipartimento di Fisica e Astronomia ``G. Galilei'', Universit\`a di Padova, I-35131 Padova, Italy}\\
$^{77}${INAF Istituto di Radioastronomia, I-40129 Bologna, Italy}\\
$^{78}${Dipartimento di Astronomia, Universit\`a di Bologna, I-40127 Bologna, Italy}\\
$^{79}${Universit\`a Telematica Pegaso, Piazza Trieste e Trento, 48, I-80132 Napoli, Italy}\\
$^{80}${Grupo de Altas Energ\'ias, Universidad Complutense de Madrid, E-28040 Madrid, Spain}\\
$^{81}${Department of Physical Sciences, Hiroshima University, Higashi-Hiroshima, Hiroshima 739-8526, Japan}\\
$^{82}${Friedrich-Alexander-Universit\"at Erlangen-N\"urnberg, Erlangen Centre for Astroparticle Physics, Erwin-Rommel-Str. 1, 91058 Erlangen, Germany}\\
$^{83}${Department of Astronomy, University of Maryland, College Park, MD 20742, USA}\\
$^{84}${Laboratoire AIM, CEA-IRFU/CNRS/Universit\'e Paris Diderot, Service d'Astrophysique, CEA Saclay, F-91191 Gif sur Yvette, France}\\
$^{85}${The George Washington University, Department of Physics, 725 21st St, NW, Washington, DC 20052, USA}\\
$^{86}${University of North Florida, Department of Physics, 1 UNF Drive, Jacksonville, FL 32224 , USA}\\
$^{87}${Science Institute, University of Iceland, IS-107 Reykjavik, Iceland}\\
$^{88}${Nordita, Royal Institute of Technology and Stockholm University, Roslagstullsbacken 23, SE-106 91 Stockholm, Sweden}\\
$^{89}${Istituto Nazionale di Fisica Nucleare, Sezione di Pisa, I-56127 Pisa, Italy}\\
$^{90}${Department of Physics, KTH Royal Institute of Technology, AlbaNova, SE-106 91 Stockholm, Sweden}\\
$^{91}${The Oskar Klein Centre for Cosmoparticle Physics, AlbaNova, SE-106 91 Stockholm, Sweden}\\
$^{92}${Dipartimento di Fisica, Universit\`a degli Studi di Perugia, I-06123 Perugia, Italy}\\
$^{93}${Hiroshima Astrophysical Science Center, Hiroshima University, Higashi-Hiroshima, Hiroshima 739-8526, Japan}\\
$^{94}${Istituto Nazionale di Fisica Nucleare, Sezione di Roma ``Tor Vergata", I-00133 Roma, Italy}\\
$^{95}${Department of Physics and Astronomy, University of Denver, Denver, CO 80208, USA}\\
$^{96}${Osservatorio Astronomico di Trieste, Istituto Nazionale di Astrofisica, I-34143 Trieste, Italy}\\
$^{97}${Funded by contract FIRB-2012-RBFR12PM1F from the Italian Ministry of Education, University and Research (MIUR)}\\
$^{98}${Department of Physics, University of Johannesburg, PO Box 524, Auckland Park 2006, South Africa}\\
$^{99}${Institut f\"ur Astro- und Teilchenphysik and Institut f\"ur Theoretische Physik, Leopold-Franzens-Universit\"at Innsbruck, A-6020 Innsbruck, Austria}\\
$^{100}${NYCB Real-Time Computing Inc., Lattingtown, NY 11560-1025, USA}\\
$^{101}${Solar-Terrestrial Environment Laboratory, Nagoya University, Nagoya 464-8601, Japan}\\
$^{102}${Max-Planck-Institut f\"ur Physik, D-80805 M\"unchen, Germany}\\
$^{103}${Institute of Space Sciences (CSICIEEC), Campus UAB, Carrer de Magrans s/n, E-08193 Barcelona, Spain}\\
$^{104}${Instituci\'o Catalana de Recerca i Estudis Avan\c{c}ats (ICREA), E-08010 Barcelona, Spain}\\
$^{105}${INAF-Istituto di Astrofisica Spaziale e Fisica Cosmica Bologna, via P. Gobetti 101, I-40129 Bologna, Italy}\\
$^{106}${Praxis Inc., Alexandria, VA 22303, resident at Naval Research Laboratory, Washington, DC 20375, USA}\\
\bsp	
\label{lastpage}
\end{document}